# Analysis of surface tension in terms of force gradient per unit area. Part II : Theoretical model of a statistical molecular reorganization gradient at interfaces

*André Schiltz*
*Grenoble, France*

## Abstract

*Classically, surface tension is seen as a force per unit length or as energy per unit area. The surface energy is calculated thermodynamically on the surface of a mathematical layer with no thickness. The surface energy concept is certainly practical and elegant, but it doesn't seem entirely satisfactory to us.*

*In an earlier article[1], we proposed a thought experiment in which surface tension is replaced with an equivalent force per unit area according to the principles of fluid mechanics. This theoretical tool can be used to rewrite the classical static equations in terms of surface stress gradient and propose a new way of calculating known phenomena such as meniscus, capillary tube, water drops, etc.*

*In this article, we propose a new thought experiment assuming that beyond the classical interaction forces acting at short distance, molecules at liquid interfaces statistically reorganize according to a stationary process of creation/destruction at long distance. The process is such that the degree of statistical molecular organization decreases with a gradient from the surface to the bulk, where the molecules recover the natural disorder of Brownian motion.*

*Assuming that this reorganization process follows an Avrami-type law, we will reinterpret the effects of surface tension in terms of energy per unit volume, which will allow us to retrieve the previous equations of force gradient per unit area.*

**Keywords :** *Surface tension, superficial tension, surface energy, capillarity, interface, potential energy, gradient.*

## I  Introduction

Surface tension is a physicochemical phenomenon related to molecular interactions of a fluid with its interface (interface with another fluid, gas, or solid). This phenomenon is responsible for the observable effects of capillarity, the formation of a meniscus at the edge of a glass, the formation of a drop of water, etc. Classically, surface tension 'γ' is defined thermodynamically as a force per unit length in [N/m] or equivalently as an energy per unit area in [J/m2].

In this thought of experience, we formulate the hypothesis of statistical reorganization at liquid interfaces according to a stationary creation/destruction process that takes place after a relatively short capillary rise time, propagates away from the interface over a millimetric rather than a nanometric distance.

This hypothesis will allow us to describe the effects of superficial tension in terms of energy per unit volume or force gradient per unit area and return to the equations of force gradient per unit area described in the previous article[1].

## II  Review of previous theories

### 2.1. Law of Laplace and equation of Young-Laplace (1804)

The concept of surface tension was defined in 1804 by Pierre-Simon Laplace[2], who calculated the pressure difference in a drop of water with respect to the atmosphere as a function of average surface curvature and surface tension 'γ' according to the so-called Young-Laplace equation :

$$\Delta P = \frac{2\,\gamma}{R} \tag{1}$$

> Where :
>> $\Delta P$ : is the pressure difference at the interface [N/m$^2$]
>> R : is the radius of the spherical drop [m]
>> $\gamma$ : is the surface tension [N/m] or [J/m$^2$]

According to this equation, the smaller the drop, the greater the pressure differential. The main limitation of the Young-Laplace equation is that the pressure theoretically tends to be infinite as the radius tends to zero. This equation is still widely used today to calculate the effects of capillarity, such as the rise of a liquid in a capillary tube according to Jurin's Law[3].

## 2.2. The Young-Dupré equation (1805)

Regarding the equilibrium of meniscus and that of a drop of water on a plane, we generally refer to Young-Dupré law[4]. According to this law, the vector sum of the following three tensions projected onto the surface is in principle zero as in the Young-Dupré equation :

$$\gamma_{LV}\cos(\theta) = \gamma_{SV} - \gamma_{SL} \tag{2}$$

> Where :
>> $\theta$ : is the contact angle of the drop or of the meniscus
>> $\gamma_{LV}$ : is the liquid-vapor surface tension
>> $\gamma_{SL}$ : is the solid-liquid surface tension
>> $\gamma_{SV}$ : is the solid-vapor surface tension

In the case of the meniscus, the diagram in **Figure 1** describes the balance of the three stresses '$\gamma_{LV}$', '$\gamma_{SL}$' and '$\gamma_{SV}$'. Note that the solid-vapor surface tensor is upward, while the liquid-vapor and solid-liquid surface tensors are downward[5].

It should be noted that most authors consider that the meniscus profile is roughly shaped like a circular arc. However, some authors are more realistic and use an exponential, such as D. Vella et al.[6] and F. Elie[7].

Interestingly, F. Elie balances the Laplace pressure with the hydrostatic pressure variation '$\rho gz$' (where '$\rho$' is the density, 'g' is the intensity of gravity and 'z' is the height of the fluid) using the general equation for the radius of curvature of the surface and writes the following differential equation : $\partial^2 z/\partial x^2 - \rho gz/\gamma = 0$.

Using the definition of capillary length : $L_c = \sqrt{\gamma/\rho g}$, he rewrites it as : $\partial^2 z/\partial x^2 - z/L_c{}^2 = 0$.

This differential equation has an exponential solution such as :

$$Z(x) = Z_0\,e^{-\,x/L_c} \tag{3}$$

According to this equation, the profile meniscus should depend on only two parameters : '$Z_0$' and '$L_c$'. The authors specify that this is an approximation.

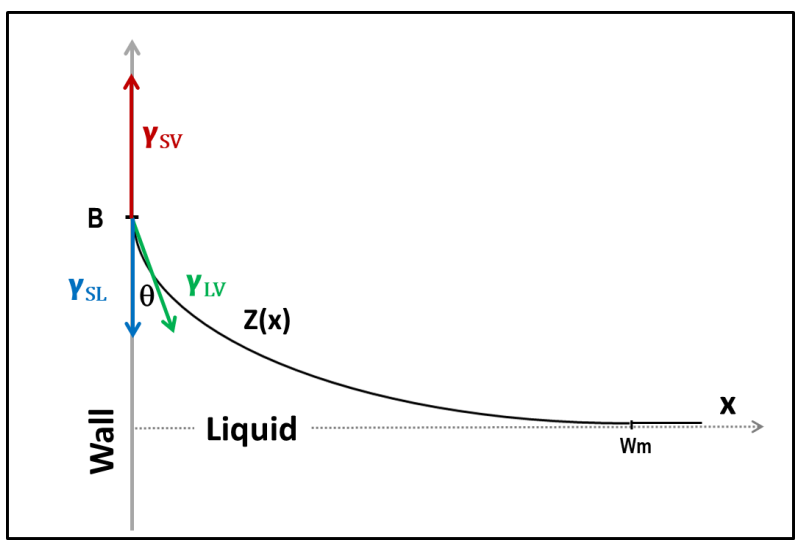

**Figure 1 – Drawing of surface tensors and contact angles for the meniscus according to the Young-Dupré equation**

### 2.3. Energy aspect : thermodynamic vision of the interface

P-G. de Genes et al.[8] consider that surface tension is a physicochemical phenomenon related to the increase in energy at the interface between two fluids or at the fluid/air interface. In a liquid, the interaction forces between molecules are in equilibrium and their resultant is zero. At the liquid-air interface, equilibrium cannot be maintained, and surface molecules are attracted inwards, creating a force known as « surface tension ». Considering that molecules at the surface have fewer interactions with their neighbors than in the volume and taking into account the Brownian effect of molecular motion[9], they estimate that if 'U' is the cohesion energy per molecule in volume, the energy at the surface should be in the order of 'U/2'.

If '$a$' is the size of a molecule, '$a^2$' is the area exposed to the surface, and it can therefore be assumed that the surface tension 'γ', which measures the loss of energy at the surface, is worth '$\gamma \approx U / 2a^2$'.

For common liquids, Van der Waals interactions predominate, and the thermal energy is calculated as ' $U \cong k_B.T$ ', where '$k_B$' is the Boltzmann constant, which gives a value of 'γ' close to *20 mJ/m²* at 25°C. In the case of water, where a surface tension three times higher is measured (γ ≈ *72 mJ/m²*), it is explained that the surface tension measured is greater because hydrogen bonds are predominant.

Thus, surface tension is thermodynamically defined as energy per unit area or, equivalently, as force per unit length. In a way, it is like considering that all the energy is concentrated at the interface on a mathematical surface with no thickness.

There are several schools of thought regarding the thickness of the molecular layer on the liquid surface : (i) J. W. Gibbs considers the surface as a mathematical surface without thickness and (ii) J. D. Van der Waals and H. Bakker assigned a thickness of the size of the Van Der Waals interactions[9].

In their 2002 bibliographic review, L. J. Michot et al.[10] confirm this vision at the molecular level, pointing out that « Whatever the support, it is now clearly evidenced that structural perturbations are limited to distances lower than 10–15 Å from the interface.» , which is 3 to 5 times the size of a water molecule.

Thus, the question of energy distribution at the interface seems to be resolved, as no structural perturbations beyond a few molecular diameters are observed.

However, if we have a look at the meniscus in a glass of water, we can see that it is not nanometric but millimetric in size (dimensions of the order of 1 to 2 millimeters in height and more than 5 millimeters in attenuation length are observed, depending on the authors[11, 12]) . The theoretical explanation is that surface forces act tangentially on the liquid surface, bending it over several millimeters and causing the liquid to move below the meniscus line.

# III  Model of superficial tension energy gradient by statistical reorganization

### 3.1 Molecular statistical reorganization hypothesis

In this thought experiment, we assume that after a relatively short period of time[(*)], a stationary reorganization process takes place beyond the nearby molecular interface, where Van der Waals type forces are at work, and propagates away from the interface over a millimeter rather than a nanometer distance. This process of statistical reorganization is the result of a mechanism whose reorganization time is probably on the order of picoseconds. Secondly, this is obviously a dynamic statistical process, not a static process. Thus, the structures formed at the interfaces have a very short lifetime. In the case of water, we can assume that these structures are formed by hydrogen bonds according to a statistical creation-destruction process. According to our hypothesis, the statistical density of this molecular reorganization varies over long distances according to a distribution gradient, from the interface to the bulk, where the molecules regain the natural disorder of Brownian motion.

So, in contrast to the classical view where all the energy is concentrated on the surface, we consider here that there is a reorganization energy per unit volume, which varies according to a gradient from the interface to the bulk. This concept results in the definition of a surface tension energy gradient per unit volume, as well as a force gradient per unit area, in contrast to the classical surface tension tensor 'γ' which represents a force per unit length.

(*) NB : The kinetics of capillary rise will not be discussed in this article.

### 3.2 Boundary conditions for meniscus

To build our model, we will use the case of the meniscus because it is ideal for several reasons :
(i) Since both the liquid and wall surfaces are large compared to the dimensions of the meniscus, they can be considered "almost infinite". (ii) There are no horizontal or vertical limitations imposed by the volume of the liquid. (iii) The meniscus profile can extend horizontally to its attenuation and climb vertically to its maximum along the solid wall as shown in **Figure 1**.

As in the Young-Dupré equation, we can therefore assume that the various forces are in equilibrium once the meniscus has formed.

More specifically, we are going to balance the gradient of superficial tension energy per unit volume against the potential energy of gravity per unit volume.

### 3.3 Construction of a reorganization model at the meniscus wall-liquid interface

First, we will ignore the influence of the solid-vapor interface and consider only liquid interfaces, i.e., the liquid-vapor interface and the solid-liquid interface. As shown in **Figure 2**, we temporarily assume that the meniscus shape results solely from the combination of surface tension forces at the solid-liquid and liquid-vapor interfaces.

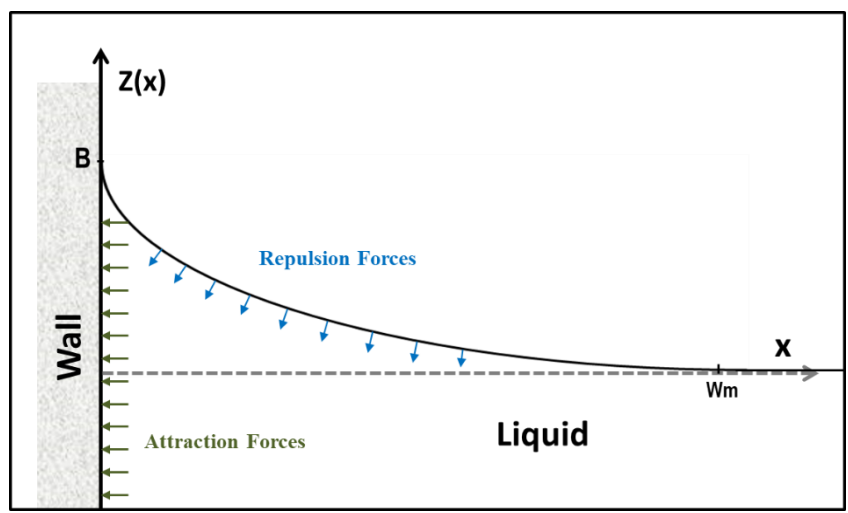

**Figure 2 – Drawing of theoretical attraction solid-liquid and repulsion liquid-air forces at meniscus interfaces**

Let us start by examining the surface tension forces at the solid-liquid interface :

### 3.3.1    Hypotheses for creating a stationary regime of molecular statistical organization

In the context of our thought experiment, to build a model of a stationary process of molecular reorganization along a distribution gradient from the interface, we need to make some prior assumptions :

- Hypothesis 1 : The energy of molecular statistical organization is of the same order of magnitude as the energy classically defined at the surface (cf. § 2.3). It is maximal at the interface and decreases along a gradient until Brownian disorder is restored at a millimeter distance. In water, for example, such organization could be achieved by building short-lived structures using hydrogen bonds in a stationary creation/destruction process that respects the principle of energy conservation.

- Hypothesis 2 : The organization process is faster than the diffusion mechanism of Brownian disorder. This hypothesis is based on the observation that the stationary equilibrium state of the meniscus is reached in a few tenths of a second, while the Brownian diffusion mechanism is relatively slow (the average statistical velocity of a water molecule is about 0.3 mm per minute, with a Brownian diffusion coefficient of about $10^{-9}$ [$m^2 s^{-1}$]).

- Hypothesis 3 : After a short dynamic formation time based on a nucleation/propagation phenomenon, a statistical organization gradient sets in under the stationary regime. In this regime, even if the lifetime of organized molecules is very short, they are quickly replaced by others through a process of creation/destruction.

- Hypothesis 4 : The organization and propagation process follows an Avrami-type law[13, 14].

On the basis of these hypotheses, we can now build a model of molecular reorganization gradient at the liquid-wall interface.

### 3.3.2    Hypothesis of creation and propagation according to an Avrami-type law

According to hypothesis 4, organization starts at the wall, where 'N' nucleation sites per unit area organize nearby liquid molecules. The organization propagates perpendicular to the interface, forming 'G' structures with a decreasing gradient from the wall to a distance 'W', where the Brownian disorder of the bulk is restored.

The phenomena of nucleation and propagation are described by the Avrami equation :

$$v = 1 - \exp(-Kt^n) \tag{4}$$

Where :

> $v$ : is the volume fraction of organized molecules
> $(1-v)$ : is the bulk molecules fraction
> $t$ : is the time
> $n$ : is a parameter depending on the degree of freedom
> $K$ : is a constant depending on $\dot{N}$ and $\dot{G}$
> $N$ : is the number of nucleation sites
> $G$ : is the degree of propagation growth

The Avrami equation is usually used to describe the kinetics of crystallization, phase changes, and chemical reactions. The 'K' constant is usually determined experimentally by measuring crystal growth and propagation rates, or chemical reaction rates.

According to our model, after a relatively short period of dynamic meniscus formation, the phenomenon becomes stationary. Growth is perpendicular to the wall and a stationary process of creation/destruction sets in. We can then take 'n = 1' and keep a simplified version of equation (4) such as :

$$v = 1 - e^{-Kt} \tag{5}$$

Where 'K' depends on the kinetic parameters of meniscus formation.

NB : Let us keep in mind that this is a thought experiment, and that no experimental measurements have yet been made to validate or invalidate the hypothesis of a statistical gradient of short-lived species in a meniscus.

### 3.3.3 Reorganization energy

According to our hypotheses, the volume fraction of organized molecules as a function of time 'v' is then given by the Avrami equation. Once stationary, their distribution in space as a function of distance from the wall is maximal at the interface and decreases along a gradient until the standard bulk disorder is reached. It is necessary to add a hypothesis about the way in which molecular exchanges take place.

- Hypothesis 5 : It is assumed that the exchange between the molecules at the interface and those in the bulk follows a law analogous to Fick's law of diffusion[15] and Fourier's law of heat exchange[16] such that :

$$d^2U/dx^2 = D\frac{\partial U}{\partial t} \tag{6}$$

Where :

> $U$: is the energy of molecular reorganization
> $t$ : is the time
> $x$ : is the distance of distribution/exchange
> $D$ : is a diffusivity coefficient

The classical solution of this equation uses the decomposition into two functions, one of distance and the other of time 'U(x, t) = f(x).g(t)' such as :

$$U(x, t) = U_0\, e^{-Kt}\, e^{-Jx} \tag{7}$$

Where :

U(x, t) : is the energy of molecular reorganization or 'surface tension energy'

$U_0$ : is its value at à x=0

K : is an Avrami-type constant that depends on the kinetic parameters of meniscus formation

J : is a spatial attenuation constant

t : is the time

x : is the energy diffusion and exchange distance

Transferring (7) to (6), we obtain : $J^2 = D\,K$, and the molecular reorganization energy can be rewritten as :

$$U(x, t) = U_0\, e^{-K\,t}\, e^{-x\,\sqrt{D\,K}} \tag{8}$$

In general, the meniscus quickly reaches a stationary state of equilibrium, i.e., after several tenths of a second[17, 18].

Reminder : We will not analyze the kinetic aspect in this article.

### 3.3.4 Stationary state assumption

In this article, we focus on the stationary state of the meniscus once it has been formed.

We can consider that after a time 't' called 'relaxation time' (t >> 1/K), a stationary state of equilibrium is reached, and we can simplify equation (8), which describes the reorganization energy as a function of distance from the surface, and rewrite it as :

$$U(x) \approx U_0\, e^{-A\,x} \tag{9}$$

Where '$\sqrt{D\,K}$' has been replaced by 'A', which here represents an attenuation constant that depends on the wall material, the liquid and the temperature. In fact, we will see later that 'A' can be more easily determined as a function of the geometric parameters of the meniscus.

### 3.3.5 Balance of organizational energy and gravitational potential energy

As mentioned above, we can assume that the various forces are in equilibrium once the meniscus has formed. So, we can balance the gradient of surface tension energy per unit volume with the potential energy of gravity per unit volume.

- Hypothesis 6 : In the gravity field and in a stationary state, the surface tension energy 'U(x)' is in equilibrium with the gravitational potential energy '$E_p(x) = m\,g\,Z(x)$'.

NB : We will come back to this assumption later, because while it is valid in the case of a meniscus where the solid wall and liquid surface are large compared to the size of the meniscus, it won't be the same for a drop of water.

We can write these energies in units of volume and define a density of gravitational potential energy '$E_{pp}$' such as :

$$E_{pp}(x) = \rho\, g\, Z(x) \tag{10}$$

Where :

> $E_{pp}$ : is the potential energy of gravity per unit volume or, more simply, the pressure gradient in [kg m$^{-1}$ s$^{-2}$], [Jm$^{-3}$] or [Nm$^{-2}$]
> $\rho$ : is the density in [kg m$^{-3}$]
> g : is the intensity of gravity in [m s$^{-2}$] or [N/kg]
> Z(x) : is the observed profile of the meniscus in [m]

Since hydrostatic pressure is generally defined as 'p = ρ g h', what we refer to here as 'potential energy of gravity per unit volume', 'Epp (x)', is in fact a pressure gradient. However, we will keep the notation 'E$_{pp}$' to avoid any confusion with global Pressure in the following calculations.

Rewriting [9] in units of volume and equating it with [10] gives :

$$\rho \, g \, Z(x) = u_0 \, e^{-A\,x} \tag{3}$$

Where 'u$_0$' is the surface tension energy per unit volume at 'x = 0'.

Along the wall, when 'x = 0', we can write 'Z(0) = B', where 'B' is the height of the meniscus at the wall and 'u$_0$' is : u$_0$ = ρ g B.
The surface tension energy gradient 'u(x)' per unit volume can be written as :

$$u(x) = \rho \, g \, B \, e^{-A\,x} \tag{4}$$

In the case of the meniscus, the final profile 'Z(x)' has the same form as that of the energy gradient and is simply written as :

$$Z(x) = B \, e^{-A\,x} \tag{15}$$

### 3.3.6 Expression of surface tension in terms of energy per unit volume or force gradient per unit surface area

Whereas surface tension 'γ' is conventionally defined as force per unit length in [N/m] or equivalently as energy per unit area in [J/m$^2$], we define here the 'superficial tension' as energy per unit volume ([Jm$^{-3}$]) or as a force gradient per unit area ([Nm$^{-2}$]).

The superficial tension energy gradient per unit volume (12) can also be written as a force gradient per unit area or stress gradient, such as :

$$\sigma_R(x) = \rho \, g \, B \, e^{-A\,x} \tag{14}$$

Where :

> $\sigma_R(x)$ : is the resulting stress gradient in [Nm$^{-2}$] or [Jm$^{-3}$]
> B : is the final meniscus height in [m]
> A : is an attenuation constant in [m$^{-1}$]
> $W_m$: is the meniscus attenuation length in [m]

This stress gradient notation '$\sigma_R$' or force gradient per unit area in [Nm$^{-2}$] will now enable us to recover the equations of the previous article[1].

# IV Merging the statistical reorganization gradient model with the mechanical vision of equivalent forces

In an earlier article[1], we proposed a thought experiment in which surface tension is replaced with an equivalent force per unit area according to the principles of fluid mechanics. This theoretical tool can be used to rewrite the classical static equations in terms of surface stress gradient and to propose a new way of calculating known phenomena such as the meniscus, the capillary tube, water drops, etc.

In this article we will phenomenologically reinterpret the previous equivalent force gradients in terms of molecular attraction or repulsion and relate them to the previous equations without re-calculations.

NB : Whenever we quote an equation from the previous article[1], we agree to write it as : (I-'ref').

### 4.1. Conventions

We have previously[1] defined the following components :

$\sigma_{SL}(x)$ : is the surface stress gradient at the solid-liquid interface.

$\tau_{LV}(x)$ : is the surface stress gradient at the liquid-vapor interface.

$\omega_{SV}(x)$ : is the surface stress gradient at the solid-vapor interface.

Concerning the meniscus, we will first analyze the action of each gradient separately, before analyzing their combined action.

### 4.2. Phenomenological analysis of the solid-liquid interface gradient

The surface tension stress gradient at the solid-liquid interface '$\sigma_{SL}(x)$' results from the attraction of liquid molecules to the wall.

According to our model, this attraction creates a gradient of statistical molecular reorganization at the solid-liquid interface, extending from the wall to a millimeter distance into the bulk. In the stationary state, this organization is constantly modified by thermal agitation and there is a gradient of statistical density of order and disorder from the wall to the bulk.

This reorganization gradient creates a stress/strain gradient parallel to the solid surface whose effect decreases from the wall to the bulk, as shown in **Figure 3**.

**Figure 3** describes several gradients : the gradient of solid-liquid attractive forces, the gradient of the theoretical statistical density of molecular organization, and the stress-strain gradient '$\sigma_{SL}(x)$'.

As indicated above, in the case of the meniscus, the forces of solid-liquid attraction and liquid-vapor repulsion are in competition. In **Figure 3**, we first have virtually represented the solid-liquid attraction component. The effect of liquid-vapor forces will be discussed later.

The expression for the attraction stress at the solid-liquid interface '$\sigma_{SL}(x)$' can be found in equation (8) in the previous article[1].

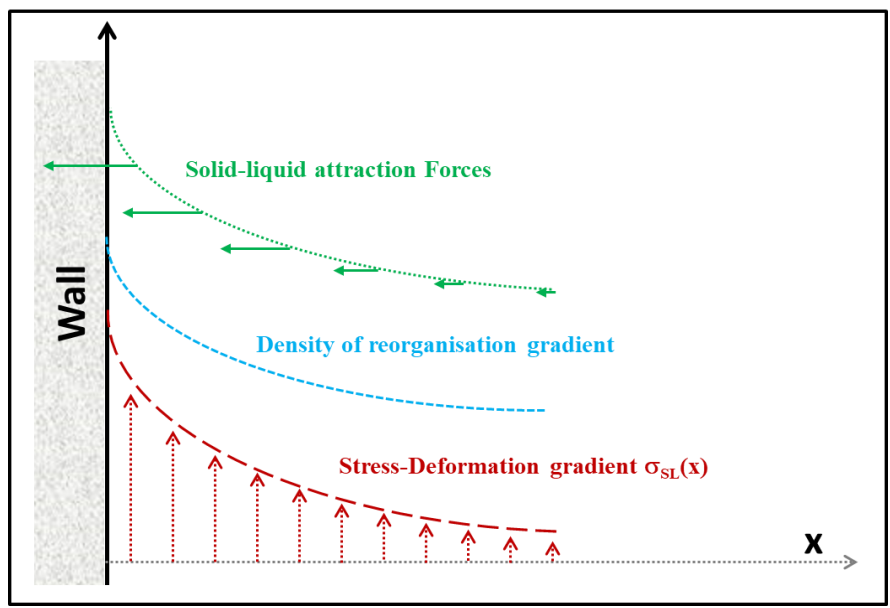

**Figure 3 - Schemes of the solid-liquid attraction forces, the theoretical reorganization gradient and the stress-deformation gradient '$\sigma_{SL}(x)$'**

### 4.3. Phenomenological analysis of the liquid-vapor interface gradient

Similarly to the solid-liquid interface, we hypothesize that molecular reorganization occurs at the liquid-vapor interface and propagates away from the surface over a millimeter depth. This time it's not an attraction of the liquid molecules towards the wall, but a repulsion of the molecules from the surface towards the bulk (equivalently, we can say as De Gennes et al.[8] do, that the surface molecules are attracted inward).

According to our model, the reorganization gradient due to the repulsion of molecules at the liquid-vapor interface creates a compression gradient '$\tau_{LV}(z)$' perpendicular to the liquid surface, the action of which decreases from the surface to the bulk.

The expression for the compression stress at the solid-liquid interface '$\tau_{LV}(x)$' can be found in equation (9) in the previous article[1].

Note that its action depends on the configuration of the liquid-vapor interface. In fact, the compression gradient '$\tau_{LV}(z)$' does not have the same effect on a large flat surface, on the meniscus, or, for example, in the case of a drop of water.

That is why we are going to analyze each of these cases.

### 4.3.1. Effect of liquid-vapor repulsion forces on a large flat surface

In the case of a large flat surface, away from the edges, the only surface tension forces involved are the liquid-vapor repulsion forces.

**Figure 4a** shows simultaneously : the liquid-vapor repulsion forces, the theoretical statistical organization density gradient, and the stress-compression gradient '$\tau_{LV}(z)$', for a large liquid surface. The liquid-vapor repulsion forces are perpendicular to the surface of the liquid and tend to compress it, but do not deform the surface because the surface is large compared to the distance at which these forces act.

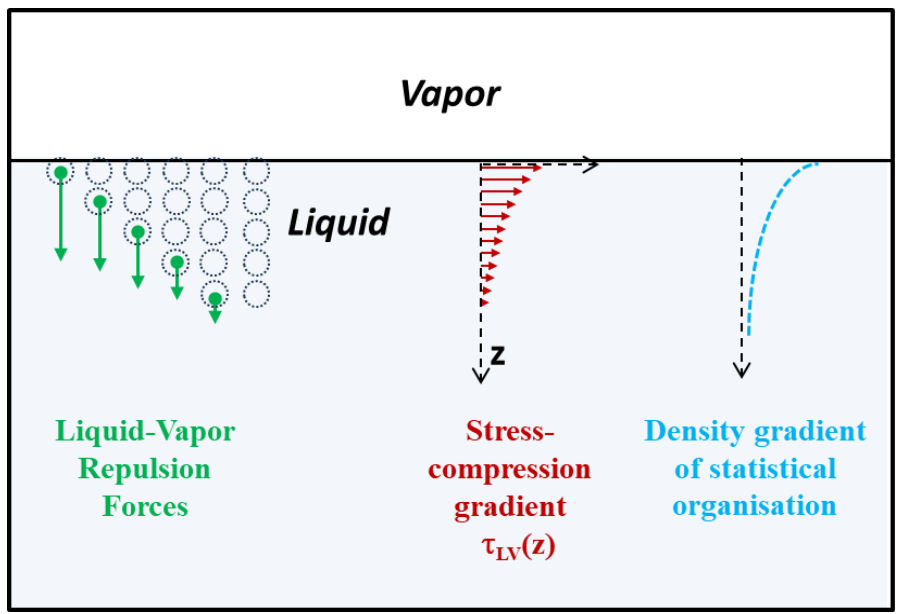

**Figure 4a – Theoretical representation of Liquid-Vapor repulsion forces, gradient of statistical organization and stress-compression gradient '$\tau_{LV}(z)$' for a large liquid surface**

### 4.3.2. Effect of liquid-vapor repulsion forces in the case of a spherical drop

In the case of a spherical drop, e.g. a drop of water, the only surface tension forces involved are the liquid-vapor repulsion forces, as in the previous case.

They are at the origin of the stress-compression gradient '$\tau_{LV}(z)$' which was described in equation (I-26) in the previous article[1].

These forces compress the drop, and since the size of the drop is on the order of the effective dimension of the surface tension forces, they tend to make it spherical.

In the previous article, we illustrated their action in **Figure I-10b**, which we have simply reproduced below.

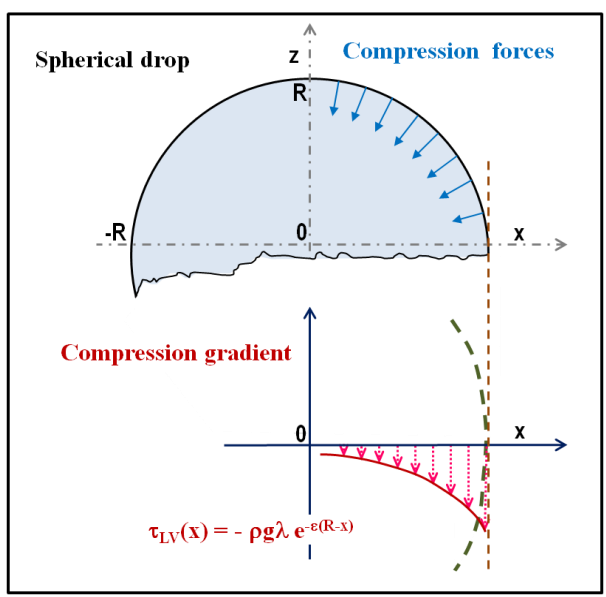

**Figure I-10b – Compression gradient '$\tau_{LV}(x)$' in a spherical drop of water due to compression forces. The stress gradient is negative and maximum at the surface, but not zero at the center.**

**Water drops size limit**

In the case of a raindrop, some observers have observed that when its radius 'R' reaches a critical radius which we will note 'R$_c$', gravity forces dominate and dislocate the drop. Indeed, drop cohesion depends on drop size, and no drop is observed beyond a diameter of 6 to 8 mm[19] (see **Figure I-10a** in the previous article).

We have previously calculated the equilibrium equation between the critical weight of the drop and the sum of the forces exerted by the compression gradient and obtained the following equation :

$$\frac{2}{3} R^2 < \lambda/\mathcal{E} \left( 1 - e^{-\mathcal{E}R} \right) \tag{6}$$

This inequation simply states that when the drop radius 'R' is smaller than 'R$_c$', the pressure difference in the drop is positive, the drop is under compression and remains compact. For a radius greater than 'R$_c$', the pressure difference is negative, and the drop is broken.

**Calculating the pressure difference in a drop and Laplace's Law**

In the previous article, we calculated the overpressure in the drop by dividing the sum of the forces acting on its perimeter '2ΠR' by the area of the median plane 'ΠR$^2$', as :

$$\Delta P = 2 \rho g \lambda \left\{ (1 - e^{-\mathcal{E}R})/(\mathcal{E}R) \right\} \tag{7}$$

The expression in square brackets is very interesting, as it tends towards unity when 'R' tends towards zero. In the previous article, we plotted equation (I-33) and Laplace (1) as a function of 'R' in **Figure I-1**. We observed that as 'R' tends to zero, the pressure in equation (1) tends to infinity, whereas in equation (I-33), pressure tends to a finite value : $\Delta P(R=0) = 2 \rho g \lambda$.

Thus, equation (I-33) solves the infinity problem of Laplace's equation when the drop radius tends to zero. Indeed, we can calculate that the pressure difference in a drop of 1 micron diameter should be about 2.8 bar according to equation (1), whereas it is only about 87.10$^{-5}$ bar according to equation (I-33), which seems more reasonable to us.

**The case of a drop of water in weightlessness**

As in the above case, the only surface tension forces involved are liquid-vapor repulsion forces which produce radial compression. However, in the case of a water drop in weightlessness, our equations suggest that we should observe water drops with a much larger radius.

As mentioned in the previous article[1], this is exactly what cosmonauts were able to observe during experiments conducted in weightlessness, where they manipulated water drops over ten centimeters in diameter[20-22].

We have assumed that the attenuation length of the exponential 'W$_g$' in weightlessness is greater than that observed on Earth in the case of the meniscus 'W$_m$' and, keeping a generic parameter as in equation (I-33), we have plotted the theoretical pressure variation in the drop beyond 'R$_c$' as in **Figure I-12**. This allowed us to extrapolate a theoretical attenuation length 'W$_g$' of around ten centimeters.

Although we do not yet have precise information on the measurement of 'W$_g$', our hypotheses remain consistent with observation.

### 4.3.3. Phenomenological analysis of the meniscus: action of liquid-vapor repulsion forces

In the meniscus case, the forces of solid-liquid attraction and liquid-vapor repulsion are in competition. Above, we have shown a virtual representation of the solid-liquid attraction component in **Figure 3**.

This time, we have represented the virtual component of the liquid-vapor repulsion forces in **Figure 4b**. The liquid-vapor repulsion forces are perpendicular to the liquid surface and do not deform it for a large liquid surface. However, in the vicinity of the wall, they should theoretically curve the surface convexly, as shown in **Figure 4a** (as opposed to the final meniscus, which is concave).

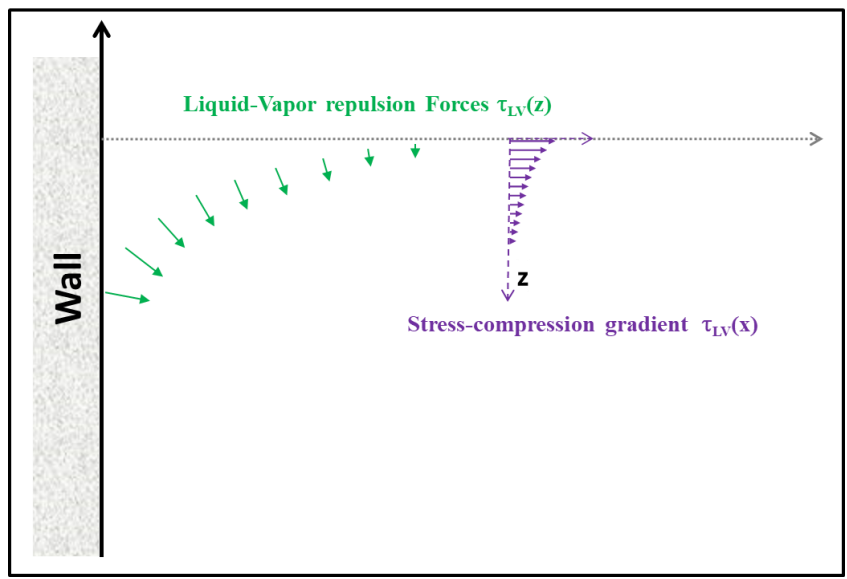

**Figure 4b – Theoretical representation of the liquid-vapor repulsive stress gradient '$\tau_{LV}(z)$' component in the meniscus case (solid-liquid attractive forces and solid-vapor forces are not considered).**

We can now analyze the combined effect of solid-liquid attraction and liquid-vapor repulsion on the meniscus.

### 4.4. Phenomenological analysis of the combined forces of solid-liquid attraction and liquid-vapor repulsion at the meniscus

The combined effect of solid-liquid attractive forces and liquid-vapor repulsive forces results in the formation of a concave meniscus as shown in **Figure 4c**. The final meniscus profile is the result of the stress gradient of deformation '$\sigma_{SL}(x)$' and compression '$\tau_{LV}(z)$'.

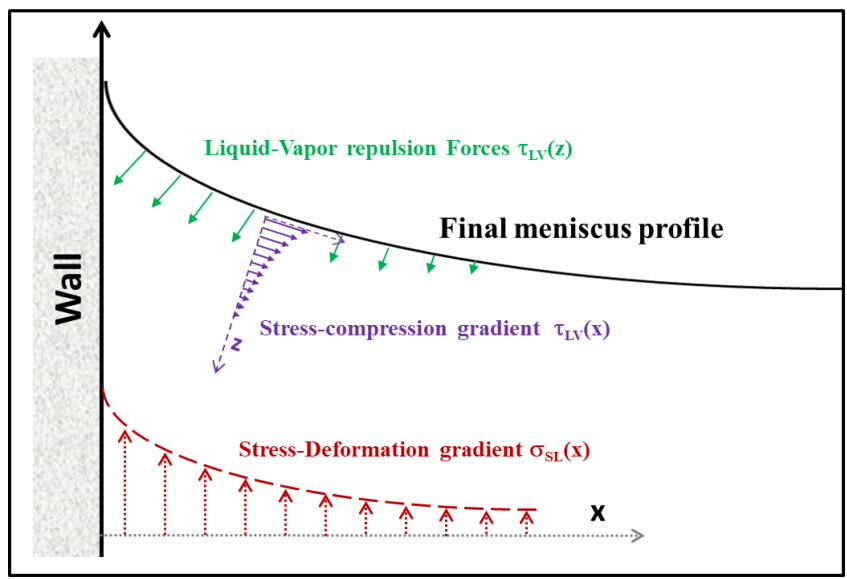

**Figure 4c – Combined action of the stress gradients '$\sigma_{SL}(x)$' and '$\tau_{LV}(z)$' in the meniscus case**

**Figure 4c** shows the normal profile of a concave meniscus. We can see that the liquid-vapor repulsion forces act on the profile created by solid-liquid attraction forces and limit the rise of the meniscus, but they do not change its overall shape.

In the profile equation, the compression gradient '$\tau_{LV}(z)$' acts perpendicular to the '$Z_{SL}(x)$' profile created by solid-liquid attraction forces. To calculate the resulting stress gradient '$\sigma_R(x)$', the stress gradient '$\sigma_{SL}(x)$' must be added to the projection of the stress gradient '$- \tau_{LV}(x)$' onto the derivative of the profile '$Z_{SL}(x)$', as written in equation (I-11)[1], which gives :

$$\sigma_R(x) = \rho g\, B\, e^{-A\,x} = \rho g\, \delta\, e^{-\alpha\,x} + \rho g\, \lambda\, e^{-\varepsilon\,x}.\, \alpha\delta\, e^{-\alpha\,x} \qquad (I-12)$$

In the case of water, the profiles calculated in this way are shown in **Figure I-5**, which we have simply reproduced below.

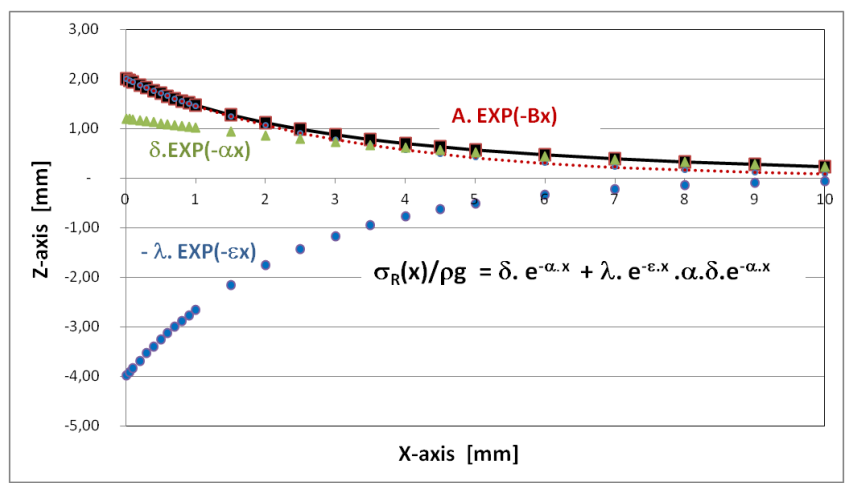

***Figure I-5 - Calculated profiles of the meniscus in a glass of water with the parameters :***
***{ B= 2.10⁻³ [m] ; A= 316 [m⁻¹] ; δ=1,2.10⁻³ [m]; α= 167 [m⁻¹] ; λ= 4.10⁻³ [m] ; ε= 409 [m⁻¹]}***

### 4.5. Molecular organization and gradient at the solid-vapor interface

Unlike what happens at solid-liquid and liquid-vapor interfaces, where molecular reorganization takes place within the liquid, molecular organization at the solid-vapor interface is achieved by condensation/ adhesion on the wall. Remember that for water, the molecular density in the gas phase is about a thousand times lower than in the liquid phase. Therefore, according to our model, the surface tension forces at the solid-vapor interface should be significantly weaker than the forces of solid-liquid attraction and liquid-vapor repulsion. As in the case of solid-liquid interaction, the organization is strongly dependent on the affinity of the vapor molecules for the solid, and the surface state of the solid is a key factor in the nucleation mechanism described in Hypothesis 4.
To account for the low density in the vapor phase, we need to add a new assumption about the process of organization.

- Hypothesis 7 : A solid-vapor molecular organization takes place on the wall above the 'triple point' by attraction and condensation of the liquid's vapor molecules.

Note that this assumption corresponds to what observers call the precursor film[23-25]. In the case of water, this is a film less than 100 nanometers thick. It extends beyond the triple point in the case of the meniscus, and beyond the drop seat in the case of a drop deposited on a substrate.
In accordance with Hypothesis 4, we can assume that there are 'N' nucleation sites per unit area on the wall, which induce an organization of nearby molecules, and that the organization spreads perpendicular to the wall, creating 'G' structures.
However, there are far fewer vapor molecules available than in the liquid phase, so the deposited film will be very thin. In addition, the film should become thinner as we move away from the triple point, because the concentration of available molecules decreases.

In the absence of information on the kinetics of condensation, attraction, and propagation of vapor molecules in the liquid, we will simply assume that the solid-vapor stress gradient '$\omega_{SV}(x)$' has the same form as the solid-liquid and liquid-vapor gradients, but not the same parameters, as described in the previous article :

$$\omega_{SV}(x) = \rho \, g \, \kappa \, e^{-\upsilon x} \tag{8}$$

Where :

$\omega_{SV}(x)$ : is the solid-vapor stress gradient in [Nm$^{-2}$] or [Jm$^{-3}$]

$\kappa$ : is a parameter related to film height in [m]

$\upsilon$ : is an attenuation constant in [m$^{-1}$]

Thus, due to the low molecular density in the gas phase, our model leads us to consider that solid-vapor interfacial forces are involved in the formation of the precursor film but play a limited role in the formation of the meniscus and drops.

Note that this view is very different from that of the Young-Dupré equation (2). It will be discussed later, in section 4.7.

### 4.6. Phenomenological analysis of combined solid-liquid, liquid-vapor, and solid-vapor forces at the meniscus

When all solid-liquid, liquid-vapor and solid-vapor surface tension components are combined, the resulting stress gradient '$\sigma_R(x)$' can be written as :

$$\sigma_R(x) = \rho \, g \, \kappa \, e^{-\upsilon x} + \rho \, g \, \delta \, e^{-\alpha x} + \rho \, g \, \lambda \, e^{-\varepsilon x} . \, \alpha \, \delta \, e^{-\alpha x} \tag{I-14}$$

In the case of water, the profiles calculated in this way are shown in **Figure I-5bis**, which we have simply reproduced below.

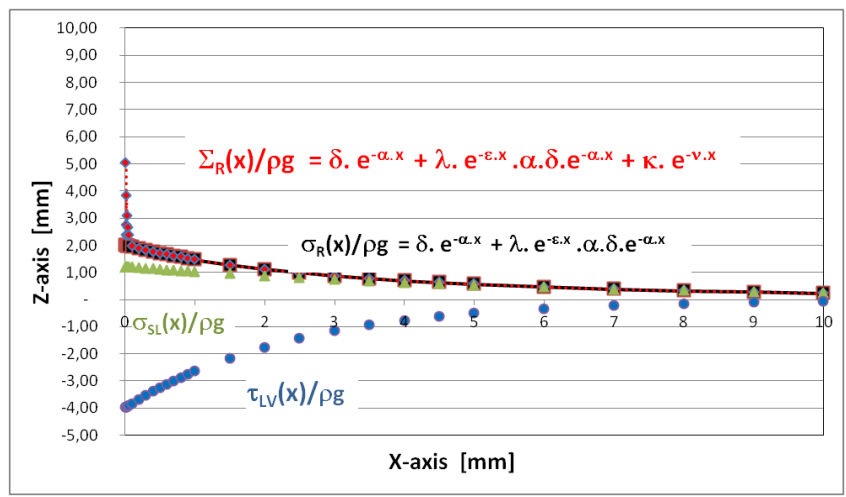

***Figure I-5bis - Calculated meniscus profiles in a glass of water with parameters : { B=2.10$^{-3}$ [m] ; A=316 [m$^{-1}$] ; δ=1,2.10$^{-3}$ [m] ; α=167 [m$^{-1}$] ; λ=4.10$^{-3}$ [m] ; ε=409 [m$^{-1}$]; κ=5.10$^{-3}$ [m]; ν=5.10$^{4}$ [m$^{-1}$] }***

According to our model, solid-vapor surface tension forces are expected to play only a very limited role in the formation of the precursor film and their inclusion has only a very slight effect on the calculated profile. In the previous article[1], we showed the existence of this precursor film in the case of the mercury meniscus in **Figure I-6a**, since it is transparent in the case of water.

### 4.7. Phenomenological analysis of surface tension forces in the case of a hemispherical drop and discrepancies with the Young-Dupré law

In the case of a hemispherical drop, the configuration is different from that of the meniscus.
According to our model, while the forces of solid-liquid attraction tend to spread the hemispherical drop parallel to the solid surface (as in the meniscus case), the forces of liquid-vapor repulsion tend to compress it and maintain its shape, as shown on the right side of **Figure I-22** from the previous article. It should be added that gravity also tends to spread the hemispherical drop.
We can see in **Figure I-22** that our interpretation of the solid-liquid and liquid-vapor interface contradicts the Young-Dupré equation (2), which assumes that the solid-vapor interface force tends to spread the hemispherical drop, while the solid-liquid and liquid-vapor interface forces tend to compress it.

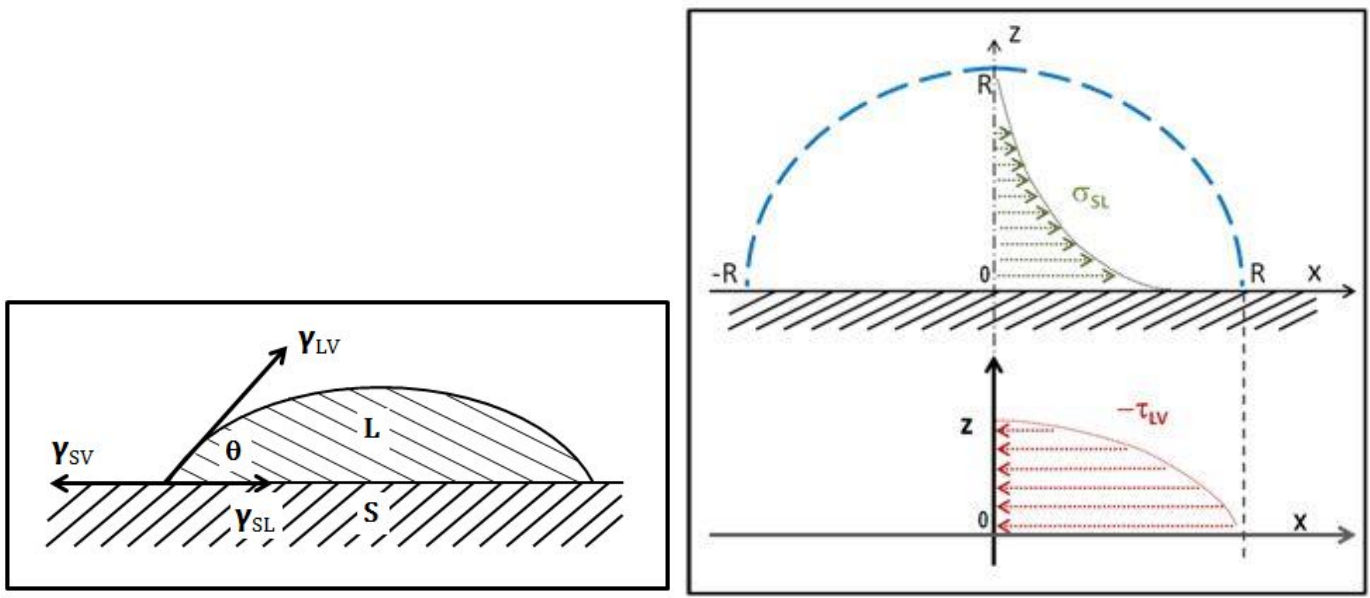

***Figure I-22 – Comparison of force vector patterns in a hemispherical drop according to the Young-Dupré equation and our model***

We can list the discrepancies between our model and Young-Dupré's law :
- The Young-Dupré equation considers that the solid-vapor interface forces play a key role in the construction of both the meniscus and the hemispherical drop.
  - In our model, we assume that the solid-vapor interface forces play only a limited role in the construction of the meniscus and droplet and that they are primarily involved in the creation of the precursor film.
- The Young-Dupré equation considers that the solid-vapor interface forces tend to spread the hemispherical drop, while solid-liquid and liquid-vapor interface forces tend to compress it.
  - We consider here that the liquid-vapor interface forces tend to compress the drop, while the solid-liquid interface forces and gravity tend to spread it out.
- The Young-Dupré equation doesn't take gravity into account, either in the case of the meniscus or in the case of the isolated drop, where gravity plays, however, an important role by limiting its size and imposing a critical radius.
  - Our model considers gravity and the critical radius.
A subsequent study should clarify some of these discrepancies.

# V DISCUSSION

In this article, we proposed a thought experiment based on the hypothesis that molecules at liquid interfaces reorganize statistically according to a stationary long-range creation/destruction process.
In the case of water, for example, such a process could take place via the creation/destruction of very short-lived structures using hydrogen bonds. Once the meniscus has formed, a stationary regime of creation/destruction could set in, independent of time but dependent on distance from the wall.
To date, no experimental measurements have been made to validate or invalidate such a model.

At the interface level, the currently prevailing hypothesis is that the structural perturbations do not extend to distances greater than 10-15 Å. This is the origin of the current theory of surface tension which considers that, in the case of the meniscus, the surface forces act tangentially on the liquid surface and stretch it over several millimeters, causing the liquid to move below the meniscus line.
In their bibliographic review, L. J. Michot et al.[10] confirm this molecular view, stating that "it is clear that, whatever the substrate, structural perturbations do not extend over distances greater than 10-15 Å", which corresponds to 3 to 5 times the size of a water molecule.
On the other hand, J.M. Zheng et al[26] disagree and report that they have observed structural disturbances at the interfaces extending over several hundred microns below the surface. These authors report that colloidal and molecular solutes suspended in an aqueous solution can be largely excluded from the vicinity of various hydrophilic surfaces over several hundred microns (« *It is generally thought that the impact of surfaces on the contiguous aqueous phase extends to a distance of no more than a few water-molecule layers. Older studies, on the other hand, suggest a more extensive impact. We report here that colloidal and molecular solutes suspended in aqueous solution are profoundly and extensively excluded from the vicinity of various hydrophilic surfaces. The width of the solute-free zone is typically several hundred microns.* »).

Until our hypotheses are validated or invalidated, for example by spectroscopic measurements at the meniscus level, our hypotheses remain a highly theoretical thought experiment.
This thought experiment has at least the merit of raising several questions about the validity of current laws in the interpretation of the precursor film, the infinity problem for nanometric drops and the size of drops in weightless conditions.

# VI CONCLUSION

In a previous article[1], we proposed a thought experiment which consists of mathematically replacing the classical tangential surface tension force with an equivalent fluid mechanics force gradient, exerted in the volume. This allowed us to reinterpret previous equations, provide an improved expression of Laplace's law, and model the formation of semi-ellipsoidal and pseudo-spherical drops.
In this article, we propose a new thought experiment assuming that molecules at liquid interfaces reorganize statistically according to a stationary long-range creation/destruction process. According to this hypothesis, the organization is such that the degree of molecular statistical organization decreases along a gradient from the surface to the bulk of the liquid, where the molecules return to the natural disorder of Brownian motion. Assuming that this reorganization process follows an Avrami-type law, we have reinterpreted the effects of surface tension in terms of energy per unit volume and recovered the force gradient equations per unit area described in the previous article.

# References

*[1] Andre Schiltz, Analysis of surface tension in terms of force gradient per unit area – Part I : a thought experiment using the principle of equivalence in fluid mechanics. 2025. DOI:10.48550/arXiv.2406.16448. (French)- hal-04572796v2.(English).*

*[2] Pierre Simon de Laplace, Traité de Mécanique Céleste, volume 4, (Paris, France : Courcier, 1805).*

*[3] James Jurin (1719) "An account of some new experiments, relating to the action of glass tubes upon water and quicksilver", Philosophical Transactions of the Royal Society of London, 30 : 1083–1096.*

*[4] Thomas Young (1805) "An essay on the cohesion of fluids," Philosophical Transactions of the Royal Society of London, 95 : 65–87.*

*[5] Marchand A, Weijs J H, Snoeijer J H and Andreotti B 2011 Why is surface tension parallel to the interface, American Journal of Physics, 999 -1008*

*[6] Dominic Vella and L. Mahadevan, The ''Cheerios effect'', Am. J. Phys. 73, 817–825 (2005)*

*[7] Frédéric Elie – Effet capillaire des liquides, https://www.researchgate.net/publication/316858944_Effets_capillaires_des_liquides_TP_sur_la_longueur _capillaire*

*[8] P. G. de Gennes, Wetting : Statics and dynamics, Rev. Modern Physics **57,** 827 (1985).*

*[9] Simone Bouquet et Jean-Paul Langeron, « INTERFACES », Encyclopædia Universalis/interfaces.*

*[10] L.J. Michot et al. : Water organisation at the solid–aqueous solution interface, C. R. Geoscience - 334 (2002) 611–631.*

*[11] Toru Shimizu - Three Kinds of Expressions for Meniscus at Flat Wall -Advanced Studies in Theoretical Physics - Vol. 14, 2020, no. 2, 73 – 79.*

*[12] Giuseppe Soligno , Marjolein Dijkstra and René van Roij - The equilibrium shape of fluid-fluid interfaces: Derivation and a new numerical method for Young's and Young-Laplace equations - J Chem Phys. 2014 Dec 28;141(24):244702.*

*[13] Melvin Avrami, J. Chem. Phys. 7, 1103 (1939); 8, 212 (1940); 9, 177 (1941)*

*[14] Humphrey J. Maris, C. R. Physique 7 (2006)*

*[15] Adolf Fick, « Über Diffusion », Annalen der Physik und Chemie, vol. 94, 1855, p. 59–86*

*[16] Joseph Fourier, Théorie analytique de la chaleur, 1822 - Édouard Leroy, « Sur l'intégration des équations de la chaleur », ASENS, 3e série, t. 14, 1897, p. 379-465*

*[17] Joachim Delannoy. Les surprises de la montée capillaire. Mécanique des fluides [physics.class-ph]. Sorbonne Université, 2019. Français. ⟨NNT : ⟩. ⟨tel-02867154v1⟩*

*[18] Giuseppe Soligno , Marjolein Dijkstra and René van Roij - The equilibrium shape of fluid-fluid interfaces: Derivation and a new numerical method for Young's and Young-Laplace equations - J Chem Phys. 2014 Dec 28;141(24):244702.*

*[19] H. R. Pruppacher and J. D. Klett, Microphysics of Clouds and Precipitation - Manfred Wendisch, 1999, Journal of Atmospheric Chemistry - J ATMOS CHEM.*

*[20] https://video-streaming.orange.fr/high-tech-science/des-astronautes-jouent-avec-de-l-eau-en-impesanteur-CNT000001zOTrn.html.*

*[21] Comportement de l'eau en apesanteur - Agence spatiale canadienne - https://www.asc-csa.gc.ca/fra/recherche/video/regarder.asp?v=1_w15r1tgh.*

*[22] Essorage à bord de l'ISS - https://www.youtube.com/watch?v=7weks82EYxg.*

*[23] M. N. Popescu, G. Oshanin, S. Dietrich and A-M. Cazabat - Precursor films in wetting phenomena – arXiv:1205.1541 [cond-mat.soft] - Journal of Physics: Condensed Matter, Volume 24, Number 24.*

*[24] Frédéric Elie - Etalement des gouttes sur une surface plane : loi de Tanner (2017)- https://www.researchgate.net/publication/316858830.*

*[25] Hossein Pirouz Kavehpour - An Interferometric Study of Spreading Liquid Films – MIT – (2003).*

*[26] J.M. Zheng et al. - Surfaces and interfacial water: evidence that hydrophilic surfaces have long-range impact - Advances in Colloid and Interface Science 127 (2006) 19–27.*